# Massively Parallel RNA Chemical Mapping with a Reduced Bias MAP-seq Protocol

Matthew Seetin[1], Wipapat Kladwang[1], J. P. Bida[1], and Rhiju Das[1,2,3*]
*Departments of Biochemistry[1], Biomedical Informatics[2] and Physics[3], Stanford University, Stanford CA 94305*

** To whom correspondence should be addressed: rhiju@stanford.edu. Phone: (650) 723-5976. Fax: (650) 723-6783.*

Running title: Massively parallel chemical mapping

## *Summary*

Chemical mapping methods probe RNA structure by revealing and leveraging correlations of a nucleotide's structural accessibility or flexibility with its reactivity to various chemical probes. Pioneering work by Lucks and colleagues has expanded this method to probe hundreds of molecules at once on an Illumina sequencing platform, obviating the use of slab gels or capillary electrophoresis on one molecule at a time. Here, we describe optimizations to this method from our lab, resulting in the MAP-seq protocol (Multiplexed Accessibility Probing read out through sequencing), version 1.0. The protocol permits the quantitative probing of thousands of RNAs at once, by several chemical modification reagents, on the time scale of a day using a table-top Illumina machine. This method and a software package MAPseeker (http://simtk.org/home/map_seeker) address several potential sources of bias, by eliminating PCR steps, improving ligation efficiencies of ssDNA adapters, and avoiding problematic heuristics in prior algorithms. We hope that the step-by-step description of MAP-seq 1.0 will help other RNA mapping laboratories to transition from electrophoretic to next-generation sequencing methods and to further reduce the turnaround time and any remaining biases of the protocol.

## *Introduction*

RNAs play an active role in all kingdoms of life (1, 2), with known functions including catalysis (2, 3), gene regulation (4), genome maintenance (5), and protein synthesis (1). The functions of RNA molecules appear to be critically dependent on their structures or, in the case of gene translation, their lack of structure. Comparative sequence analysis, x-ray crystallography, and NMR (6, 7) remain the gold standards for inferring functional secondary structures and tertiary interactions in RNA molecules. However, all three of these tools require significant investigator insight and effort and are not generally tractable for all RNAs. When these methods have not been applicable, chemical probing has been a valuable tool, particularly to falsify or support preexisting structural models (8-10). Enhancing the information content of chemical probing to permit *de novo* structure determination, to leverage novel chemical probes, and to accelerate RNA design are ongoing areas of research (11-15). Most chemical probes tend to react with nucleotides that are solvent-exposed, flexible, or otherwise uninvolved in canonical or non-canonical base pairing (16-22). These reactivities are often read out via a reverse transcriptase reaction that extends a DNA primer complementary to the modified RNA (16-22). The reverse transcriptase stops at the site of a modification, leaving a complementary DNA (cDNA) whose length marks the modified nucleotide. To identify the sites of the reactive nucleotides, the pool of cDNAs is typically analyzed through length separation by gel or capillary electrophoresis (CE), using radiolabeled or fluorophore-labeled primers, respectively.

With magnetic bead purifications and 96-well plates, mapping several dozen RNAs with a set of different chemical reagents is now feasible on the time scale of a day (see Chapter on the Mutate-and-Map method, in this volume). However, some applications, such as characterizing RNAs discovered by high-throughput biological screens or created by computational design (13), require probing 1000s of RNAs. Enzymatic mapping, which gives sparse reads, can be carried out on such large pools—or, indeed, at a genome-wide scale (23-25)—through next-generation-sequencing, but these data do not carry the information content of single-nucleotide-resolution chemical probes. An experimental protocol and bioinformatics pipeline developed by Lucks and colleagues has pioneered the use of 2´-OH acylating reagents and Illumina next-generation sequencing (NGS) to enable highly parallel chemical mapping of several RNAs (26, 27), but the method has not yet been applied to RNAs of many distinct sequences or with alternative chemical probes (26). Furthermore, in our hands, this method leads to mapping profiles that do not agree within error with measurements from replicate mapping experiments based on capillary electrophoresis. The differences are at least partially due to the sequence-dependence of ligation and amplification steps that have been used to prepare libraries for Illumina sequencing (unpublished data).

To increase the rate at which a large number of RNAs may be probed and to reduce bias, we have further advanced this protocol into a MAP-seq (Multiplexed Accessibility Probing) method (Figure 1). Multiple chemical reactions, such as DMS alkylation (12), carbodiimide modification of bases (12, 28) and acylation of the 2´-hydroxyl (21, 22) can be carried out in parallel (Figure 2). Addition of an Illumina adapter sequence to reverse transcribed DNA is carried out under conditions that give near-quantitative single stranded DNA ligation (29, 30). Unique sequence identifiers at the 3´ ends of each the RNA are sequestered into hairpins (Figure 1, Table 1) to help prevent interference with the fold of the RNA segment of interest. Solution PCR steps, which can introduce significant length and sequence bias, are eliminated. Last, we noticed that current NGS sequence analysis packages (31-33) either became unusably slow and/or inaccurate in assigning reads at the 3´ ends of sequences and across highly similar sequences (e.g., single mutants in mutate-and-map experiments). We have therefore developed a new automated tool, called MAPseeker, which quantifies the reactivities of the nucleotides of each RNA under each condition.

Comparisons of this NGS approach and prior CE methods are shown for different chemical modifiers in Figure 2. The data from the two techniques agree within the estimated error of each method, as can be seen in direct overlays for two RNAs (Figure 3). These data suggest that MAP-seq biases have been reduced to the point that the NGS-based data are as trustworthy as CE data.

The MAP-seq protocol is rapidly evolving, with active developments to enhance multiplexing through "index" reads in Illumina machines, to probe oligonucleotide pools and sub-pools derived from massively parallel synthesis technologies, and to tackle RNAs with lengths of kilobases. Continuing improvements in single-stranded DNA ligation, through new enzymes and solution conditions (34, 35), may further reduce any residual bias at this step. This chapter gives a snapshot of our stable "version 1.0" lab protocol that runs on the table-top MiSeq platform for RNA pools with lengths up to 200 nucleotides.

## *Materials*

### General Materials

The protocol given below assumes that four chemical probes will be tested on the pool of RNAs: no modification (addition of water or DMSO), DMS, CMCT, and 1M7. We carry out some steps of this procedure within wells of a 96-well plate, which enables use of magnetic separators and multichannel pipetters and can reduce the experimentalist's effort. Alternatively, the entire protocol can also be carried out in RNAse-free non-stick Eppendorf tubes. We use a capillary electrophoresis (CE) sequencer and fluorescent primers to check the yield and length distribution at two steps prior to Illumina sequencing. Use of fluorescent DNA is particularly convenient for rapidly piloting improvements to the protocol and for directly comparing CE data to Illumina data. As an alternative, steps to monitor the lengths of DNA fragments can also be carried out without fluorescent primers on, e.g., Agilent Bioanalyzer systems (26).

1. One 96-well V-bottom plate (E&K scientific).
2. Three or more plastic seal films (EdgeBio Systems).
3. 24-post and 96-post magnetic stand (VP Scientific).
4. Plenty of RNase-free sterile water.
5. 10 mL 70% ethanol, freshly prepared.
6. At least 120 µL 40% polyethylene glycol 8000.
7. ROX-Formamide mix: Mix 1 µL of ROX 350 ladder (Applied Biosystems) in 350 µL of Hi-Di formamide. This volume is enough for five 4-sample reactions and can be stored frozen at –20 °C.
8. Two 96-well format dry incubators (Hybex, SciGene), set to 55 °C and 90 °C before starting the chemical mapping part of this protocol. If these types of incubators are not available, a water bath or thermocycler may be substituted.
9. Access to an ABI 3100 sequencer or similar equipment for capillary electrophoresis.

### Chemical Modification of RNA

It is assumed that the investigator has already obtained the RNA molecules of interest, either by chemical synthesis, *in vitro* transcription, or extraction from cells or viruses. The protocol below has been tested most thoroughly on pools of RNAs that have identical primer binding sites (PBS) at their 3´ ends. In addition, it is critical that the last segment of each sequence (RID in Figure 1), immediately 5´ to the reverse transcription binding site, be unique in the pool (26, 27). These nucleotides (8 in our example cases below; see Table 1) are used later to identify the RNA with which each NGS read is associated. In our examples, these 8 nucleotides are arbitrary octamers, but we also included additional complementary nucleotides to sequester the RID into a hairpin capped by a UUCG tetraloop (Table 1), which reduces the potential for this added sequence interfering with the RNA fold. Aside from these PBS and RID segments, there are few other constraints. We have carried out MAP-seq on RNA pools with between one and 4000 distinct sequences in a single experiment.

Different chemical modification reagents probe different nucleotides and different aspects of the RNA structure. Each reagent may need its own quench. We list the recipes for most common modification reagents and their quenches at concentrations and reaction times appropriate for RNAs under 100 nucleotides in length. Longer RNAs should use lower concentrations and/or shorter reaction times. When deviating from these conditions or using alternative reagents, the RNA molecules should be modified with single-hit frequency, which may depend on reagent concentration, reaction time, and temperature. The protocol given below assumes that four chemical probes will be tested on the pool of RNAs: no modification (addition of water or DMSO), DMS, CMCT, and 1M7. Every 4-sample reaction will require the following:

1. Each reaction should use 2.5 to 5.0 pmol of total RNA (5 μL of 0.5–1.0 μM RNA).
2. 10.0 μL of 0.5 M Na-HEPES, pH 8.0 (from a sterile stock filtered with a 0.2 μm filter). Other pH stocks can be used, but note that carbodiimide and 2´-hydroxyl modification reactions, as well as other modifers, are sensitive to pH.
3. 20 μL of 50 mM $MgCl_2$ if the RNA requires magnesium for proper folding. If not, 20 μL 5 M NaCl can be used to give 1 M NaCl conditions; or other folding buffer.
4. 280 μL Agencort AMPure XP (Beckman Coulter) for RNA purification. (*See* **Note 1** *and* **Note 2**).
5. DMS mix for dimethyl sulfate mapping, which attacks exposed Watson-Crick faces of adenines and cytosines in RNA and DNA: Mix 1.0 μL fresh dimethyl sulfate (Sigma) into 9.0 μL 100% ethanol, then add 90.0 μL of sterile water. The quench for this reaction is 2-mercaptoethanol (Sigma). Each DMS reaction needs 5 μL DMS mix.
6. CMCT mix for carbodiimide mapping using 1-cyclohexyl-(2-morpholinoethyl)carbodiimide metho-p-toluene sulfonate (CMCT), which attacks exposed Watson-Crick faces of guanines and uracils in RNA: Mix 42 mg/ml of CMCT in $H_2O$. The solid stock for reagent is usually kept at -20 ° C. Before mixing into water, incubate it at room temperature for 15 minutes, immediately prior to experiment. The quench for this reaction is 0.4 M Na-MES, pH 6.0. Each CMCT reaction needs 5 μL CMCT mix.
7. 1M7 mix for 2´-hydroxyl acylation mapping using 1-methyl-7-nitroisatoic anhydride (1M7), a fast-acting reagent which attacks dynamic nucleotides by modifying the 2´ hydroxyl of an RNA: Mix 10 mg of 1M7 into 1 mL anhydrous DMSO. The quench for this reaction is 2-mercaptoethanol (Sigma). Each 1M7 reaction needs 5 μL 1M7 mix. An alternative reagent for 2´-hydroxyl acylation is N-methylisatoic anhydride (NMIA).

**Reverse Transcription**

1. 4 μL 5X First Strand buffer (Invitrogen)
2. 1 μL 0.1 M DTT (Invitrogen)
3. 1.6 μL dNTPs mixture (10 mM each nucleotide) per sample plate
4. 1.0 μL 40% PEG 8000
5. 0.4 μL 200 U/μL SuperScript III (Invitrogen) reverse transcriptase
6. 0.5 μL each of at least four 1 μM reverse transcription primers per sample. These primers contain the Illumina adapter for bridge amplification at the 5´ ends, followed by the Illumina sequencing primer binding sequence, followed by a primer identifier sequence (PID; see Figure 1), and finished with a region to bind the conserved 3´ ends of the RNAs. The PID region is 12 nucleotides long and must be sequence-balanced as part of a set of four primers (*See* **Note 3** *and* **Table 1**). The different PIDs can be used to distinguish between the same RNA modified by different reagents or subject to different modification reaction conditions. The primers should be purified by polyacrylamide gel electrophoresis (PAGE) or HPLC.

7. 0.25 μL of 1 μM of at least one of the reverse transcription primers with a 5´ fluorescein (FAM) modification.
8. Streptavidin-coated magnetic beads that have bound double-biotinylated oligos complementary to Illumina's Oligo C sequence (Oligo C´ beads). These beads are assembled using DynaBeads (Invitrogen), using the manufacturer's instructions for coupling to oligonucleotides under RNAse-free conditions. The oligonucleotide sequence ordered from Integrated DNA Technologies is in **Table 1**.
9. 40 μL 0.4 M NaOH
10. Acid quench mix, prepared by mixing 1 volume of 5 M NaCl, 1 volume of 2 M HCl, and 1.5 volumes of 3 M sodium acetate (pH 5.2). This stock can be kept at room temperature for at least two months. Of this mix, 40 μL are needed per 4-sample reaction.

**Ligation of Second Illumina Adapter**

Every 4-sample reaction requires:

1. 2.5 μL 10 μM Read 2 PBS/Amplification 2 (**Fig. 1, Table 1**). This contains standard sequences used for bridge PCR amplification and 2$^{nd}$ read sequencing on Illumina platforms.
2. 2 μL CircLigase ssDNA Ligase (Epicentre, Illumina)
3. 2 μL CircLigase ssDNA Ligase buffer (Epicentre, Illumina, supplied with CircLigase)
4. 1 μL 1 mM ATP (supplied with CircLigase)
5. 1 μL 50 mM $MnCl_2$ (supplied with CircLigase)
6. 2 μL 40% polyethylene glycol 8000
7. 1 μL fluorescent oligo loading control mix: 1 nM fluorescein-labeled 20-nucleotide DNA oligomer and 1 nM fluorescein-labeled 40-nucleotide DNA oligomer (*See* **Note 4**).

**Final Preparation for Sequencing**

For every 4-sample reaction, you will need

1. 0.5 μL 10 nM PhiX genomic DNA (20 nM of single strands), prepared for sequencing (Illumina, *see* **Note 9**).
2. 2 μL 2.0 N NaOH (Illumina, or other supplier)
3. 10.0 μL EB buffer (Qiagen)

## *Methods*

All steps are at room temperature unless specified.

**Chemical Modification**

1. Prepare modification buffer at 2x concentration. For each 20 μL reaction, mix 2 μL 0.5 M Na-HEPES, pH 8.0, 4 μL appropriate salt solution (e.g., either 50 mM $MgCl_2$ or 5 M NaCl, as above), and 4 μL RNase-free $H_2O$. For four reactions, prepare enough premix for five reactions (e.g., 10 μL 0.5 M Na-HEPES, pH 8.0; 20 μL of 50 mM $MgCl_2$; and 20 μL RNase-free $H_2O$).
2. Add 5 μL of each RNA solution to as many samples in the 96-well plates as you have reagents and reaction conditions. Add 10 μL of modification buffer to each well. Seal the plate with sealer tape. Let the RNAs fold at room temperature for 20 minutes, or, if necessary, place into a thermocycler, and take the plate through the necessary temperature steps to ensure proper folding of the RNA or RNAs in question.
3. While the RNA is folding, prepare fresh stocks of modification reagents as above. Always make sure that one of the modification conditions is just $H_2O$ or just anhydrous DMSO, as a control reading for background reverse transcriptase stops. Array the various modification reagents in the auxiliary plate so that they may be easily added to the main reaction plate.
4. Add 5 μL of the modification reagent to each corresponding well of the main reaction plate. For four modification conditions [(A) no modification, (B) DMS, (C) CMCT, and (D) 1M7], use 5 μL of each stock reagents freshly prepared as above. The total volume in each of four reactions is now 20 μL.
5. Seal the plate and let incubate. For RNAs under 100 nucleotides with the modification reagents listed above, 30 minutes is appropriate. For longer RNAs, less time may be necessary so as to achieve single-hit kinetics. Alternative reagents may require their own modifications to this protocol (*See* **Note 5**).
6. While the RNAs are incubating with the modification reagents, array the quenches in corresponding locations in the auxiliary plate (or in a new auxiliary plate) so that they may be easily added to the main reaction plate. After the appropriate incubation time, add 20 μL of the corresponding quench to each well. The volume of each sample is now 40 μL.
7. Prepare the bead purification mix: for each reaction well: mix 70 μL AMPure XP beads with 30 μL 40% polyethylene glycol 8000. Pipette up and down at least 10 times to mix. Add 100 μL bead purification mix (2.5x volumes) to each reaction. Again pipette up and down at least 10 times to mix. Wait 10 minutes for the beads to bind the RNA (*See* **Note 2**).
8. Set the plate on a 24-post magnetic stand and wait 5 minutes for the beads to collect near magnets.
9. Pipette off the supernatant and discard.
10. Rinse the beads with 100 μL 70% ethanol. Shake briefly with the plate on the stand(s). Wait 30 seconds. Remove the 70% ethanol supernatant
11. Repeat step 10. This time, take extra care to remove the last traces of the 70% ethanol supernatant with a P10 pipette (*See* **Note 6**).
12. Let the samples dry for 10 minutes while remaining on the magnetic stand. Remove any final visible droplets of 70% ethanol with a P10 pipette.
13. Resuspend the beads in each well in 2.5 μL RNase-free $H_2O$.Take the plate off the magnetic stand(s). Let stand for 5 minutes to ensure release of RNAs.
14. Place the plate back on the magnetic stand(s). Wait 1-2 minutes for the beads to collect. Transfer the supernatant in each well, which now contain the released, modified RNA, to a new plate.

**Reverse Transcription**

1. Prepare fresh primer stocks (*See* **Note 4**), 1 μM each. Primers should be used in sets of 4, which will be referred to as A-D within one set—primer A for the first reaction in a set of four, primer B for the second, etc. All primers in a set need to be used, due to sequence balancing requirements of the Illumina platform (*See* **Note 3, Table 1** for example sequences). For at least one primer (primer A in this example), however, instead of using pure unlabeled primer, prepare a mixture from equal parts 1 µM unlabeled primer A and 1 µM fluorescently-labeled primer A to use as the primer for the reaction in a group of four. This fluorescent tracer will allow for a rapid quality check by standard capillary sequencing prior to running on the Illumina sequencer. It is also fine to use pure fluorescently labeled primer A, or to label all primers in use. The fluorophore labeled cDNA does not interfere with the Illumina sequencing protocol (it is washed away during the bridge PCR on the Illumina flow cell).

2. Add 0.5 µL 1.0 µM of the corresponding primer to appropriate well of the plate containing the modified RNAs.
3. Prepare the SuperScript III reverse transcription mix: for each reaction, mix 1 µL of 5X First Strand buffer, 0.25 µL of 0.1 M DTT, 0.4 µL of 10 mM dNTPs, 0.25 µL 40% PEG 8000, and 0.1 µL Superscript III reverse transcriptase.
4. Add 2.0 µL of the SuperScript III reverse transcription mix: for each reaction mix to each well. Mix well by pipetting up and down at least 10 times. The total volume of the reactions is 5 µL.
5. Seal the sample plate well. Put it in an incubator set to 55 ºC for 30 minutes. For longer RNAs, longer incubation times may be used to help complete reverse transcription.
6. After 30 min, hydrolyze the leftover RNA by adding 5 µL of 0.4 M NaOH to each well. Incubate the plate at 90 ºC for 3 minutes.
7. Add 5 µL of the acid quench mix (see recipe in Reverse Transcription materials section) to each well. Let the plate cool on ice for 5 minutes.
8. Add 4 µL of the Oligo C´ beads (see recipe in Reverse Transcription materials section) to each well. Mix thoroughly by pipetting up and down at least 10 times. Incubate at room temperature 10 minutes to ensure cDNA binding to the beads.
9. Place the plate on a magnetic stand and wait 1-5 minutes for the beads to collect.
10. Pipette off the supernatant and discard.
11. Rinse the beads with 100 µL 70% ethanol. Shake briefly by hand with the plate on the stand(s). Wait 30 seconds. Remove the 70% ethanol supernatant.
12. Repeat step 10. This time, take extra care to remove the last traces of the 70% ethanol supernatant with a P10 pipette (*See* **Note 6**).
13. Let the samples dry for 5 to 10 minutes while remaining on the magnetic stand. Remove any final visible droplets of 70% ethanol with a P10 pipette, and let dry further for 5 minutes.
14. Take the plate off the magnetic stand(s). Resuspend the beads in each well in 2.5 µL RNase-free $H_2O$.
15. Take 0.5 µL from every sample reverse transcribed with the primer A/fluorescent primer A mixture and transfer to a new plate. Add 10 µL of the ROX-formamide mixture to each of these representative samples and mix. Wait 5 minutes to ensure release of cDNA.
16. Place this new plate on a magnetic stand. Wait 1-5 minutes for the beads to aggregate. Transfer the supernatant to an optical plate used by an ABI capillary electrophoresis sequencer. These samples can be loaded on an ABI sequencer to ensure reverse transcription worked properly before proceeding to the ligation step (see below); alternatively they can be loaded at the same time as the ligated samples to save effort.
17. The remaining samples (still with beads) may be pooled for convenience before proceeding. Ensure the beads are still well-suspended in water by pipetting up and down.

In the example case with one pool of RNAs modified under 4 conditions, the total volume of the combined samples will be 8 μL.

**Ligation of the Second Illumina Adapter**

1. Set one incubator or water bath to 68 ºC. Set a second at 80 ºC.
2. Prepare the CircLigase ligation mix: for every 4-sample reaction, combine 2.5 μL of 10 μM DNA oligo of second Illumina adapter sequence, phosphorylated at both the 5´ and 3´ ends; 2.0 μL CircLigase ssDNA Ligase buffer; 1.0 μL 1 mM ATP; 1.0 μL 50 mM $MnCl_2$; 2.0 μL 40% polyethylene glycol 1500; 1.5 μL of RNAse-free $H_2O$; and 2.0 μL CircLigase ssDNA Ligase (*See* **Note 8**). The adapter can also be blocked at its 3´ end with a spacer modification instead of a phosphate.
3. Add 12 μL of CircLigase ligation mix to the combined cDNA sample pool of each 4-sample reaction. Incubate at 68 ºC for 2 hours, then 80 ºC for 10 minutes to inactivate the ligase. This incubation can be carried out in a thermocycler. The oligo C´ beads do not appear to affect the efficiency of the ligation (unpublished data).
4. Add 5 μL 5 M NaCl and mix thoroughly by pipetting up and down. This will ensure proper resuspension of the beads, and raise the salt concentration to ensure proper hybridization of cDNA to oligo C´ beads.
5. Remove 10 μL to a well of a new 96-well plate. This sample fraction will be another quality check prior to sequencing. The remaining 15 μL of the pool will be used for sequencing.
6. Wait 5 minutes to ensure rebinding of the cDNAs to the beads. Place the plate on magnetic stand(s). Wait 1-5 minutes for the beads to collect.
7. Pipette off the supernatant and discard.
8. Rinse the beads with 100 μL 70% ethanol. Shake briefly with the plate on the stand(s). Wait 30 seconds. Remove the 70% ethanol supernatant.
9. Repeat step 8. This time, take extra care to remove the last traces of the 70% ethanol supernatant with a P10 pipette (*See* **Note 6**).
10. Let the samples dry for 5 minutes while remaining on the magnetic stand. Remove any final visible droplets of 70% ethanol with a P10 pipette.
11. For the sequencing samples, take the plate off the magnetic stand. Resuspend the beads of a pool in 8 μL of RNase-free $H_2O$ per 4 samples in the pool. Leave these pools separate until after quantifying the amount of cDNA in each pool.
12. For the quality check samples, take the plate off the magnetic stand. Resuspend each pool of beads in 10 μL of ROX-Formamide mix. Place the plate back on a magnetic stand. Wait 1-5 minutes for the beads to collect. Transfer the supernatant to an optical plate appropriate for an ABI capillary electrophoresis sequencer. Add 1 μL fluorescent oligo loading control mix (*See* **Note 4**) and mix well by pipetting up and down. Run these samples on the ABI capillary electrophoresis sequencer.

**Quality Check of Ligation and Quantification of Final Sample for Sequencing**

The following steps are based on using the suite of HiTRACE scripts for MATLAB (https://simtk.org/home/hitrace) (36). The general strategy for this step is to integrate the area of the peaks corresponding to the two fluorescent oligos in the loading control mix, which each should correspond to approximately 1 fmol of sample, average them, and then compare their areas to the combined area under the peaks corresponding to the ligated samples. Below are commands to be executed within MATLAB after having downloaded the HiTRACE scripts and added them to the MATLAB path. MATLAB and other computer syntax, functions, and variable names are written in an alternate font.

1. Verify that the ligation proceeded correctly. Execute the following command:
   ```
   d = quick_look({'foldername_with_ABI_data'}, lowerbound, upperbound)
   ```
   `foldername_with_ABI_data` corresponds to the path of the folder containing the data from the quality check immediately after reverse transcription, and the quality check after ligation. The numbers `lowerbound` and `upperbound` depend on the length of the RNA(s) in question. In the resulting view of the capillary traces, ensure that bands are visible for the reverse transcribed product, typically with a strongest band for the fully extended product (see Figure 4). Another band may be visible for unextended primer, although in many experiments we do not see this peak (Figure 4A), as our primer (Table 1) tends to be efficiently extended. In the second, ligated sample, check that this pattern of peaks has shifted to longer lengths (later elution times) after ligation (Figure 4B). In principle this ligation should be quantitative but this is not always the case, and the fraction ligated should be estimated (**see Note 7**).
2. Subtract the baseline from the data contained in d and store it in a matrix called d_bsub:
   ```
   d_bsub = baseline_subtract_v2(d);
   ```
3. Plot the baseline-subtracted data:
   ```
   figure(5) % This creates a new figure window within MATLAB
   in which to plot the data.
   plot(d_bsub(:,2))
   ```
   Here, it is assumed that the second column of `d_bsub` corresponds to the ligated sample. Use this plot to identify the bounds on the time axis for the shorter oligo peak, the longer oligo peak, and the fully-ligated cDNA region. Record these values as `shorter_lower`, `shorter_upper`, `longer_lower`, `longer_upper`, `cdna_lower`, and `cdna_upper`.
4. Calculate the area of the peaks in the cDNA region by summation:
   ```
   cdna_area = sum(d_sub(cdna_lower:cdna_upper,2))
   ```
5. Calculate the area of the peak for the shorter oligo:
   ```
   shorter = sum(d_bsub(shorter_lower:shorter_upper,2))
   ```
6. Calculate the area of the peak for the longer oligo:
   ```
   longer = sum(d_bsub(longer_lower:longer_upper,2))
   ```
7. Calculate the number of fmols in the 15 μL of sample that can be loaded on the Illumina platform (*See* **Note 9**)
   ```
   fmols = 24 * cdna_area / (shorter + longer)
   ```

**Final Preparation for Sequencing on a MiSeq**

1. These instructions, especially the amounts of cDNA loaded, are specific for running on a MiSeq using Illumina's version 2 kits. If an alternative platform is used, this protocol will need to be appropriately adjusted. Place a MiSeq version 2 reagent kit in a pool of water as deep as the fill line on the kit to thaw. Place the HT1 buffer vial in the kit on the bench top to thaw, then keep on ice.
2. The target is to load 30 fmol of cDNA onto the MiSeq plus 5 fmol of the PhiX genome as a positive control. Thirty divided by the number of pools of samples is the number of femtomoles taken from each pool. Given the quantifications of the number of femtomoles in each pool calculated above, calculate the volume of sample needed to be taken from each pool and contributed to the final pool that will be used for sequencing.

3. If the yield of cDNA from a pool is high, such that the volume of sample to be taken from the pool is less than 0.5 μL in order for it to provide its amount of cDNA, then first take a small aliquot of beads and add EB buffer to dilute down such that the target number of femtomoles will be in a 0.5 to 4.5 μL volume.
4. If the yield of cDNA from a pool is low, such that the volume of the library needed to get the desired amount of cDNA is in excess of 4.5 μL, put the pool in a well on a 96-well plate, and place the plate on a magnetic stand. Wait 1-2 minutes for the beads to collect. Discard the supernatant. Remove the plate from the stand. Resuspend the beads in the desired volume of EB buffer. Our experiments have show that the beads do not release a significant amount of cDNA into the supernatant when resuspended in water at the end of the ligation stage. Instead, NaOH is necessary to denature the DNA and release it from the beads (see steps 7 and 8 below).
5. If the volume of the final combined pool of the cDNA library is less than 4.5 μL, add EB buffer such that the final volume equals 4.5 μL.
6. Take 1.0 μL of 10 nM PhiX control sample and put it in a well of a 96-well plate. Add 1.0 μL of EB buffer. Mix by pipetting up and down. Take 0.5 μL of this solution and add it to the final pool of cDNA (on beads).
7. Take 2.0 μL of 2.0 M NaOH and put it in a well of a 96-well plate. Add 18 μL of deionized water and mix by pipetting up and down. Add 5 μL of this solution to the final pool of cDNA beads plus PhiX. Mix thoroughly by pipetting up and down. The final volume will be 10 μL.
8. Let the solution stand for 5 minutes for the DNA to denature. Place the plate on a magnetic stand. Wait 1-5 minutes for the beads to collect. Transfer the supernatant to a 1.5 mL tube.
9. Add 990 μL of chilled HT1 buffer to the 1.5 mL tube. Mix thoroughly by pipetting up and down. Transfer 375 μL of this solution to a new 1.5 mL tube. Add 225 μL of additional HT1 buffer and again mix thoroughly by pipetting up and down. (*See* **Note 10** *and* **Note 11**).
10. Proceed following Illumina's protocol for loading this final 600 μL solution in the reagent cartridge and loading the cartridge and flow-cell on the MiSeq. Use paired-end sequencing so as to be able read out both the position of the reverse transcription stop and the identity of the RNA in the cluster. The RNA can be of arbitrary length as each sequence is identified by its 3´ RID (**Fig. 1**) during the first read on Illumina's 50-cycle kits. The first read must be long enough to read through the primer's PID, the 20-nucleotide RNA binding site (RBS), and then through enough nucleotides of the RID region to distinguish the sequence from any other in the pool. Our RNAs typically have RID regions of 8-15 nucleotides. Such an RNA pool would require a first read length of 47 nucleotides. The second read must be long enough to uniquely identify the position of the reverse transcription stop compared to any other position in the RNA. We typically use 20-30 nucleotides, although this may vary depending on the sequence of the RNA. If RNAs in the pool contain long repetitive sequences, the second read will need to be longer so as to extend into the non-repetitive regions (The 50 cycle kit can, when used in the manner of this protocol, be used to read 76 nucleotides, summed over the first and second read.) Illumina's kits include read allotments for indexing above and beyond the stated read lengths that we do not use – we use the barcoding and identification schemes described here instead.

**MAPseeker analysis**

1. The MAPseeker package is freely available for download at: [https://simtk.org/home/map_seeker](https://simtk.org/home/map_seeker). Follow the instructions in the README to install on a Mac or other Unix-based machine.
2. Prepare two text files in FASTA format, with a label for each sequence on one line preceded by a ">" character, and the sequence following on the next line. The first FASTA file ("`RNA_sequences.fasta`") should contain the full-length sequences of every RNA in the run. The second ("`primers.fasta`") should contain the full sequence of every reverse transcription primer used in the run.
3. The MAPseeker executable takes input in the following format:
   ```
   $ MAPseeker -1 read1.fastq  -2 read2.fastq  -l
   RNA_sequences.fasta  -p primers.fasta  -n 8
   ```
   The files `read1.fastq` and `read2.fastq` are the FASTQ format sequence files from the MiSeq. The two .fasta files are those prepared in step 1. The flag "`-n`" specifies the length of the RID region of the RNA (8, in this example).
4. As output, MAPseeker gives summary statistics: the numbers of sequences found in the FASTQ files, the number in which the RNA binding site was found, the number that contained a primer ID that was specified in `primers.fasta`, and the number of sequences from read 1 and read 2 that matched an RNA sequence specified in `RNA_sequences.fasta`. It also gives text files named stats_ID1.txt through `stats_IDN.txt`, where *N* is the number of primers listed in `primers.fasta` (stats_ID1.txt corresponds to the first primer in primers.fasta, etc.). The text files contain tab-separated matrices of numbers. Each row corresponds to an RNA as they were listed in the `RNA_sequences.fasta` file. Each column corresponds to the site where reverse transcription stopped, beginning at 0 (full extension of the RT primer back to the RNA 5´ end) and ending at the 3´ end. Stops in the RID region as specified by "`-n`" above will all be zero, as they cannot be attributed conclusively to any one RNA. The values correspond to the number of clusters from the sequencing run of the particular RNA that stopped at that position.
5. The data in these text files are tab-delimited, so they are amenable to analysis with a most visualization or plotting programs, or to be further processed with custom analysis scripts. MAPseeker includes scripts in MATLAB (located in the src/matlab/ subdirectory) for quick data analysis and visualization. This protocol will describe the use of these MATLAB scripts. For ease of use, add the map-seeker/src/matlab/ directory to your MATLAB path, and set the present working directory to the directory containing the stats_ID*N*.txt files as well as the two input FASTA files. Run from within MATLAB:
   ```
   [D, RNA_info, primer_info, D_correct, D_correct_err ] =
   quick_look_MAPseeker('RNA_sequences.fasta','primers.fasta',
   './')
   ```
   This gives several windows. One is a histogram of counts per primer, and counts per RNA (figure window 1 within MATLAB; see Fig. 5A). Further windows give visualization of the counts (with estimated errors) for the four most highly represented

RNAs and 'reactivities', corrected for reverse transcriptase attenuation as given by the following equation (26, 37):

$$R(\text{site } i) = F(\text{site } i) / [F(\text{site } 0) + F(\text{site } 1) + \ \ldots\ + F(\text{site } i)] \qquad \text{(eq. 1)}$$

If the ligation of the fully extended product [$F$(site 0)] was not quantitative, this factor can be taken into account in the `quick_look_MAPseeker` function (**Note 7**). The 1D profiles of the RNAs with the most reads are shown (figure windows 2 and 3 within MATLAB; see Fig. 5B), as well as 2D representation of the entire data set (Figure windows 4 and 5 within MATLAB; see Fig. 5C). The 2D representations permit visual confirmation that the ID region of the RNA is folded into a hairpin; the 5´ strand of the stem should be protected to chemical modification, and the loop should be reactive. (The chemical reactivity of the 3´strand of the hairpin is not observable as it is used to identify the RNA; stops in this region result in reads that cannot be uniquely aligned to the library of RNA sequences.) All figure windows are also automatically saved to disk as encapsulated PostScript (EPS) files.

6. The output matrix D contains the raw counts of stop positions for each RNA. The matrices `D_correct` and `D_correct_err` have the reactivities and their estimated errors. They are cells with one matrix for each primer. `RNA_info` has information on the sequences and descriptions of the RNA library. The errors are based on Poisson statistics (note: sites with zero counts are assumed to have errors of ±1). To perform background subtraction of the data for primer 2 (a DMS condition in this example), using the control data for primer 1 (a mock treatment condition in this example) execute:
```
[D_sub, D_sub_err] = subtract_data( D_correct{2},
D_correct{1},D_correct_err{2}, D_correct_err{1} );
```
Repeat this for any other pairs of modification reagent condition and its corresponding mock treatment condition to generate similar matrices.
7. These data may be output into an RDAT file for convenient sharing and analysis. The necessary RDATkit scripts come installed with HiTRACE, or a separate installation can be downloaded at https://simtk.org/home/rdatkit. Then type:
```
filename = 'Example_output.rdat';
name = 'My Library';
annotations = {'experimentType:StandardState','chemical:Na-
HEPES:50mM(pH8.0)','chemical:MgCl:10mM','temperature:24C','
processing:overmodificationCorrectionRigorous','processing:
backgroundSubtraction','modifier:DMS};
comments = {'Test of Miseq data output.', 'Used MAPseeker
v1.0'};
output_MAPseeker_data_to_rdat_file( filename, name, D_sub,
D_sub_err, annotations, comments, RNA_info );
```
The RDAT file is formatted with the specifications available at http://rmdb.stanford.edu/repository/specs/. Structure analysis scripts, including helix-by-helix confidence estimation, are available on-line at: http://rmdb.stanford.edu/structureserver/. RDAT files can also be read and further analyzed via Python or MATLAB with the RDATkit scripts (https://simtk.org/home/rdatkit).

## *Notes*

1. Alternative purification. If the number of wells used ends up being small (enough to fit in a microcentrafuge) or if the length of the RNA is less than approximately 50 nucleotides (under the cutoff for AMPure XP purification), ethanol precipitation (0.1 volumes of 3 M sodium-acetate pH 5.2, then 3 volumes of cold ethanol; incubation in dry ice for 20 minutes; spin >15000g in a microcentrifuge for 20 minutes) can be used instead of AMPure XP purification. However, AMPure XP beads are more convenient for purifying a large number of samples.
2. AMPure XP beads should not be mixed with the modification reagents for longer than the times prescribed in this protocol. The reagents and quenches may damage them and ruin the purification. If the beads turn from brown to green, then the purification has likely failed.
3. Illumina's MiSeq platform requires that the first 12 reads made by the sequencer are "sequence-balanced," namely, that there are approximately an equal number of A's, C's, G's, and T's read across all the clusters. This is not much of a problem for genomic DNA, but the cDNA libraries described here all begin with the same 20-nucleotide primer binding site, which would ruin the sequence balancing. So, immediately upstream of the primer region is a 12-nucleotide sequence-balancing region. We also use this region as a unique primer identifier (PID) for which chemical modification reaction a given cluster came from. Primers must be made and used in sets of four; within each set of four, one primer must have an A at the first position, another must have a C at the first position, etc. Our sequences (**Table 1**) were designed to minimize secondary structure with the primer region (as assessed with NUPACK) (38) while maintaining sequence balance and disallowing 4 or more Gs in a row in any primer.
4. The loading control sequences we use are in Table 1. We had these sequences on hand for other applications. Other sequences should work just as well, as long as their lengths are less than the reverse transcription primers used above, so that their signals appear at locations distinct from the library in the capillary electrophoretic trace. Note that at low 1 nM concentration, these oligos may stick to the sides of conventional 1.5 mL tubes, lowering the accuracy of the control over time; use of non-stick tubes and storage of stocks at concentrations greater than 100 nM are recommended.
5. Only one site of modification can be identified by reverse transcription, so it is important to choose reaction conditions that modify each RNA once or not at all. Longer RNAs have more sites that can be modified, so it is important to lower reaction times and/or concentrations of reagents.
6. Ethanol is strong inhibitor of reverse transcription and ligation. It needs to be completely removed by pipetting and evaporation, with no remaining liquid visible and with the beads appearing dried out.
7. The factor of 24 comes from the fact that the number of femtomoles of sample being read by the sequencer is 1/8 of the amount of primer used in the transcription, and the remaining sample is 1.5 times larger than the aliquot used in the quantification. The final factor of two comes from averaging the areas of the smaller and larger oligo peaks, by which the area under the library peaks is divided.
8. Using CircLigase (i.e., TS2126 thermostable RNA ligase), we typically observe that the ligation of the adapter to reverse transcription products is near quantitative (>80%) in a sequence-independent manner, within the detection limits of capillary electrophoresis. The one exception is that the fully extended product is ligated with poorer efficiency (~50%; Figure 4A and 4B); it is not yet clear if this is due to a sequence bias (our transcripts begin with GG, leaving CC at the 3´ ends of these cDNA primer extension products) or to addition of some ligation-blocking modification by SuperScript III reverse transcriptase on cDNA that has completely extended to the end of an RNA. In any case, if the ligation efficiency of the full product is known, it can be specified at the time of data analysis and correction as a fourth argument in quick_look_MAPseeker. The default correction factor is 2.0; changing this number will affect the overall scale of the final reactivities and, less sensitively, the correction for attenuation which reweights the balance of short and long reverse transcription products [equation (1)].

9. Accurate quantification of the cDNA library is critical for acquiring high-quality sequencing data. Loading too much sample will ruin the results due to too many overlapping clusters of DNA that cannot be distinguished from one another by the sequencer, and loading too little may not yield enough clusters for the Illumina platform to find the flow cell during initial auto-focus stages. Also note that double-stranded DNA, such as the PhiX genomic DNA we recommend to be co-loaded with the single-stranded cDNA library, produces twice as many clusters per mole as single-stranded cDNA.
10. While it is possible to design a protocol that could conceivably discard less material than this one, the volumes involved for the early steps may be so small, depending on the yield after ligation, that it would typically be necessary to dilute down some of the pools of beads after ligation, and then end up discarding the leftover anyway.
11. The 625 μL of solution left over after this step should be discarded, unless it is going to be used in a parallel MiSeq sequencing run on the same day. Our experience is that the solution is not useable for additional runs even if stored overnight at 4 ºC.

## *Acknowledgements*

We thank members of the Das laboratory for discussions and tests of the protocol. We thank Tom Mann and Frank Cochran for preparing the 1M7 acylating reagent, and S. Mortimer for an updated 1M7 synthesis protocol. Writing was supported by the NIH (IRACDA fellowship to MS; R01GM100953 to RD) and the Burroughs-Wellcome Foundation (Career Award to RD).

**Table 1. Sequences used in this protocol, all shown from 5´ to 3´.** Color coding corresponds to regions highlighted in Fig. 1. Abbreviations for modifications are those used by Integrated DNA Technologies: /5phos/, 5´-phosphate; /3phos/, 3´-phosphate; /iBioT/, internal biotinylated T; /3Bio/, 3´ biotin; /5-6FAM/, 5´-6-fluorescein.

| | |
|---|---|
| **Example DNA Transcription Templates** | TTCTAATACGACTCACTATAGGAAAATATTAATTCTTTAATAAAAACTATCCGTTCGCGGATAGAAAAGAAACAACAACAACAAC<br>TTCTAATACGACTCACTATAGGAAACAAAAAAAAAAACGGCGATACGGATCGAGGCGAATTCGCCACATAGAAATATGGCGGAAAAAACAGACGGTTCGCCGTCTGAAAAGAAACAACAACAACAAC<br>TTCTAATACGACTCACTATAGGAAACAAAAAAAAAAACGGCGATACGGATCGAGGCGAATTCGCCACATAGAAATATGGCGGAAAAAAATGCACGTTCGCGTGCATAAAAGAAACAACAACAACAAC<br>TTCTAATACGACTCACTATAGGAAACAAAAAAAAAGGGCAGATCGAAAGATCAGTATAAGGGTACAGTACGAGGGTACAGCAAAAAAAACGCGACTTCGGTCGCGTAAAAGAAACAACAACAACAAC<br>TTCTAATACGACTCACTATAGGAAACAAAAAAAAAGGGCAGATCGAAAGATCAGTATAAGGGTACAGTACGAGGGTACAGCAAAAAAAGGTTAGCTTCGGCTAACCAAAAGAAACAACAACAACAAC<br>TTCTAATACGACTCACTATAGGAAACAAAAAAAAACTGCACGAAAAAATTCGAGCTGAAAACAGCAGGTTAAAAAGCCAGCCAAAAAACATAGCGTTCGCGCTATGAAAAGAAACAACAACAACAAC |
| **Example RNAs** | GGAAAAUAUUAAUUCUUUAAUAAAAACUAUCCGUUCGCGGAUAGAAAAGAAACAACAACAACAAC<br>GGAAACAAAAAAAAAAACGGCGAUACGGAUCGAGGCGAAUUCGCCACAUAGAAAUAUGGCGGAAAAAACAGACGGUUCGCCGUCUGAAAAGAAACAACAACAACAAC<br>GGAAACAAAAAAAAAAACGGCGAUACGGAUCGAGGCGAAUUCGCCACAUAGAAAUAUGGCGGAAAAAAAUGCACGUUCGCGUGCAUAAAAGAAACAACAACAACAAC<br>GGAAACAAAAAAAAAGGGCAGAUCGAAAGAUCAGUAUAAGGGUACAGUACGAGGGUACAGCAAAAAAAACGCGACUUCGGUCGCGUAAAAGAAACAACAACAACAAC<br>GGAAACAAAAAAAAAGGGCAGAUCGAAAGAUCAGUAUAAGGGUACAGUACGAGGGUACAGCAAAAAAAGGUUAGCUUCGGCUAACCAAAAGAAACAACAACAACAAC<br>GGAAACAAAAAAAAACUGCACGAAAAAAUUCGAGCUGAAAACAGCAGGUUAAAAAGCCAGCCAAAAAACAUAGCGUUCGCGCUAUGAAAAGAAACAACAACAACAAC |
| **Reverse Transcription Primers A-D** | AATGATACGGCGACCACCGAGATCTACACTCTTTCCCTACACGACGCTCTTCCGATCTACCAGGCGCTGGGTTGTTGTTGTTGTTTCTTT<br>AATGATACGGCGACCACCGAGATCTACACTCTTTCCCTACACGACGCTCTTCCGATCTGAGGCCTTGGCCGTTGTTGTTGTTGTTTCTTT<br>AATGATACGGCGACCACCGAGATCTACACTCTTTCCCTACACGACGCTCTTCCGATCTCTTTAAAATATAGTTGTTGTTGTTGTTTCTTT<br>AATGATACGGCGACCACCGAGATCTACACTCTTTCCCTACACGACGCTCTTCCGATCTTGACTTGCACATGTTGTTGTTGTTGTTTCTTT |
| **Illumina Adapter for cDNA ligation** | /5phos/AGATCGGAAGAGCGGTTCAGCAGGAATGCCGAGACCGATCTCGTATGCCGTCTTCTGCTTG/3phos/ |
| **Oligo C´ [double biotinylated]** | TGTGTAGATCTCGGTGGTCGCCGTATCATTTTTTTTTTTT/iBiodT//3Bio/ |
| **Loading Control 1** | /5-6FAM/AAAAAAAAAAAAAAAAAAAAGTTGTTGTTGTTGTTTCTTT |
| **Loading Control 2** | /5-6FAM/GTTGTTGTTGTTGTTTCTTT |

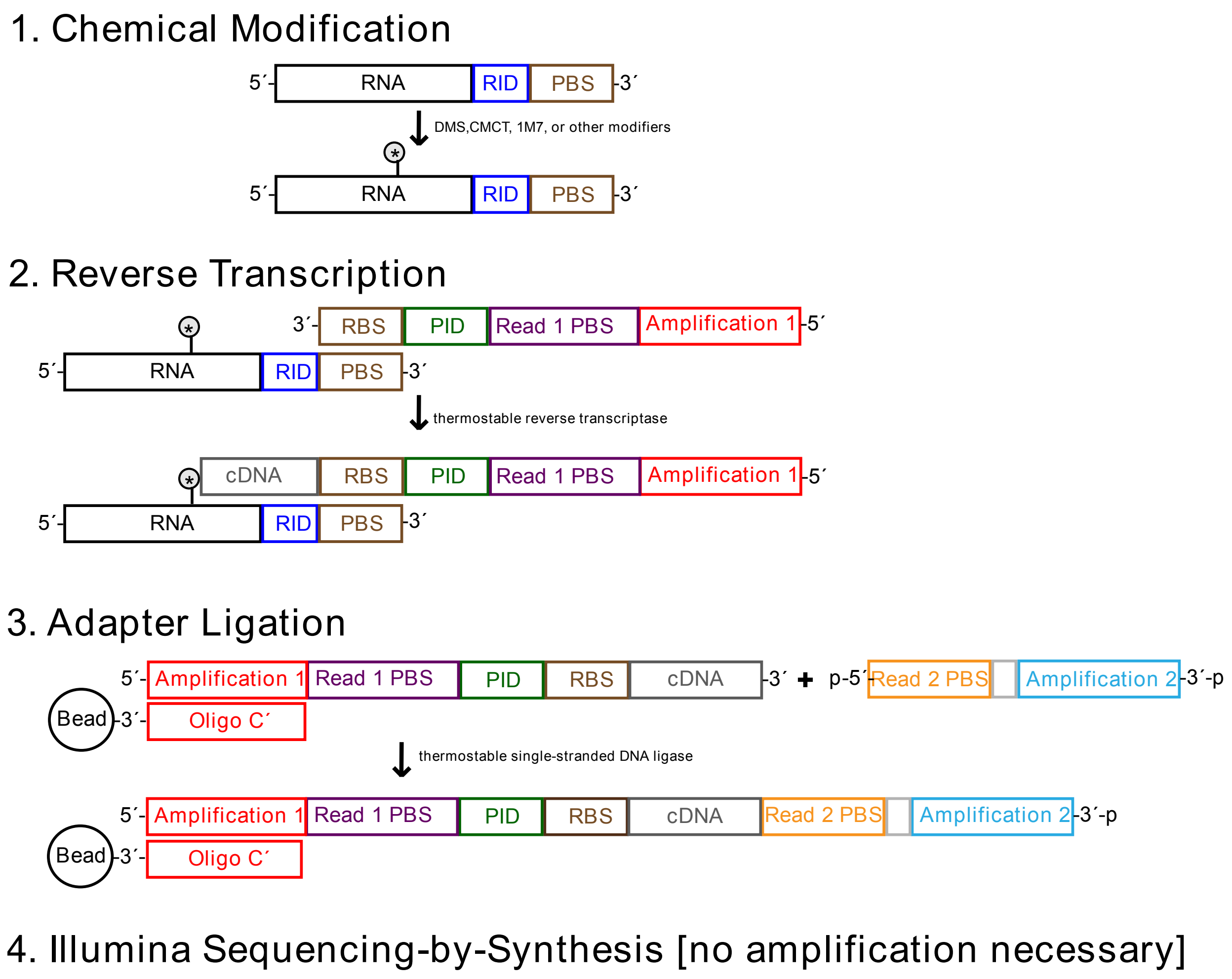

**Figure 1.** Overview of the MAP-seq experimental protocol. Each RNA has a primer binding site at its 3´ end (PBS), preceded by an RNA ID (RID) unique to each sequence, which we usually sequester into a hairpin (Table 1). First, the RNA is subject to modification by a chosen chemical reagent. The site of modification is denoted with an asterisk. Second, the RNA is reverse transcribed using a primer that has at its 3´ end an RNA binding site (RBS) complementary to the primer binding site (PBS). The primer contains a sequence-balanced 12-nucleotide primer identification region (PID), and segments necessary for sequencing read 1 and bridge amplification in the Illumina flow cell. Third, the second Illumina adapter containing the read 2 PBS, and the other DNA necessary for bridge amplification is ligated to the 3´ end of each cDNA with a single-stranded DNA ligase. Additional sequence segments within the Illumina adapters (gray) can be used for further multiplexing of experimental conditions, but are not used in the current protocol.

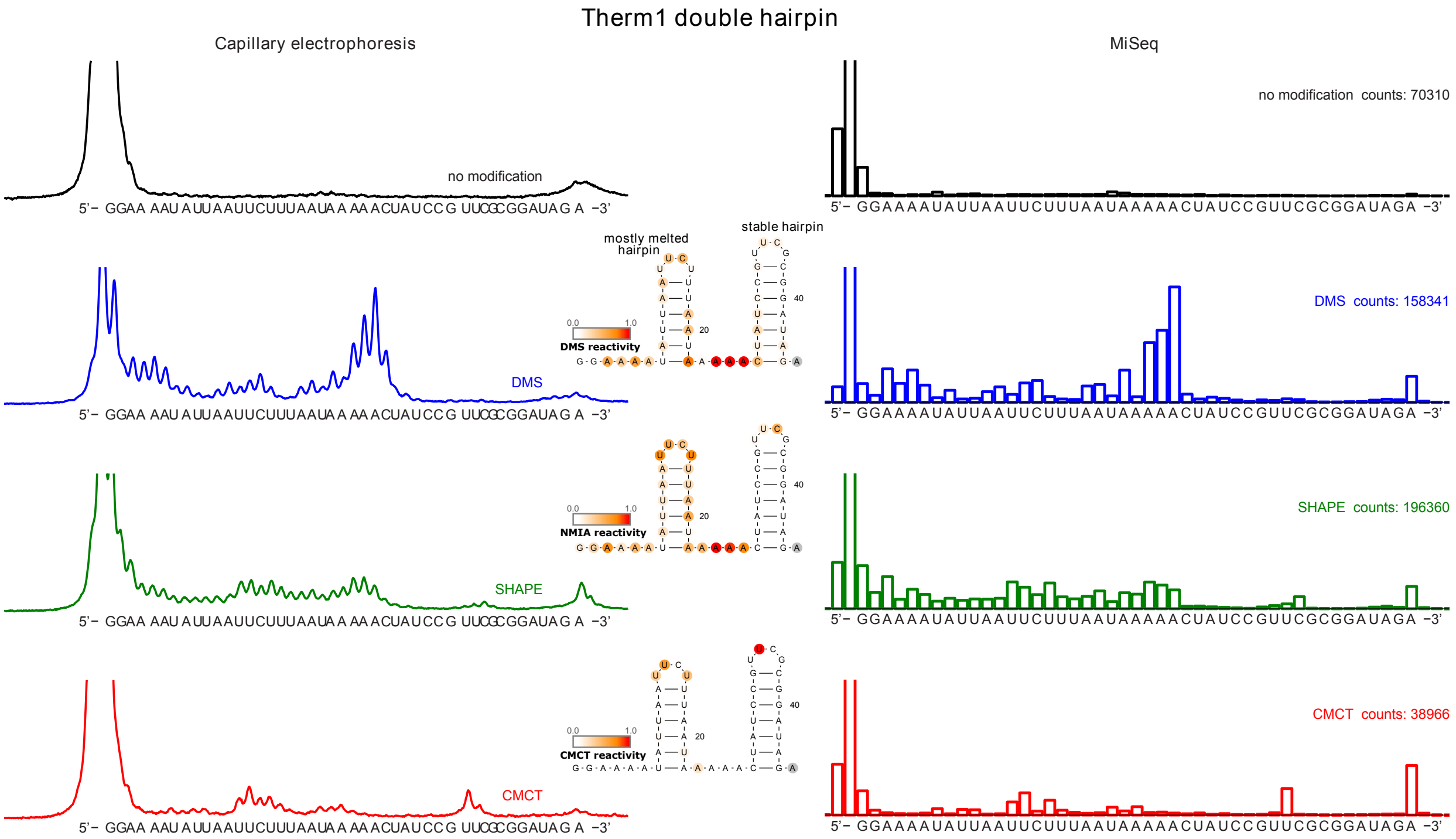

**Figure 2.** Comparison of capillary electrophoresis data (CE) and MAP-seq data on a double hairpin RNA Therm1, probed under control conditions (no modifier), with dimethyl sulfate (DMS), N-methylisatoic anhydride (NMIA), and 1-cyclohexyl(2-morpholinoethyl) carbodiimide metho-p-toluene sulfonate (CMCT). Raw data are shown: experimental capillary electrophoretic traces (left, arbitrary units; longer products on left) and number of reads (right). When quantitated and displayed on the predicted secondary structure (middle; red indicates reactivity of most reactive nucleotide), the modifiers indicate that the first of two predicted hairpins in the Therm1 construct is significantly melted in the tested conditions (24 °C; 1 M NaCl; 50 mM Na-HEPES, pH 8.0), as compared to the second longer hairpin.

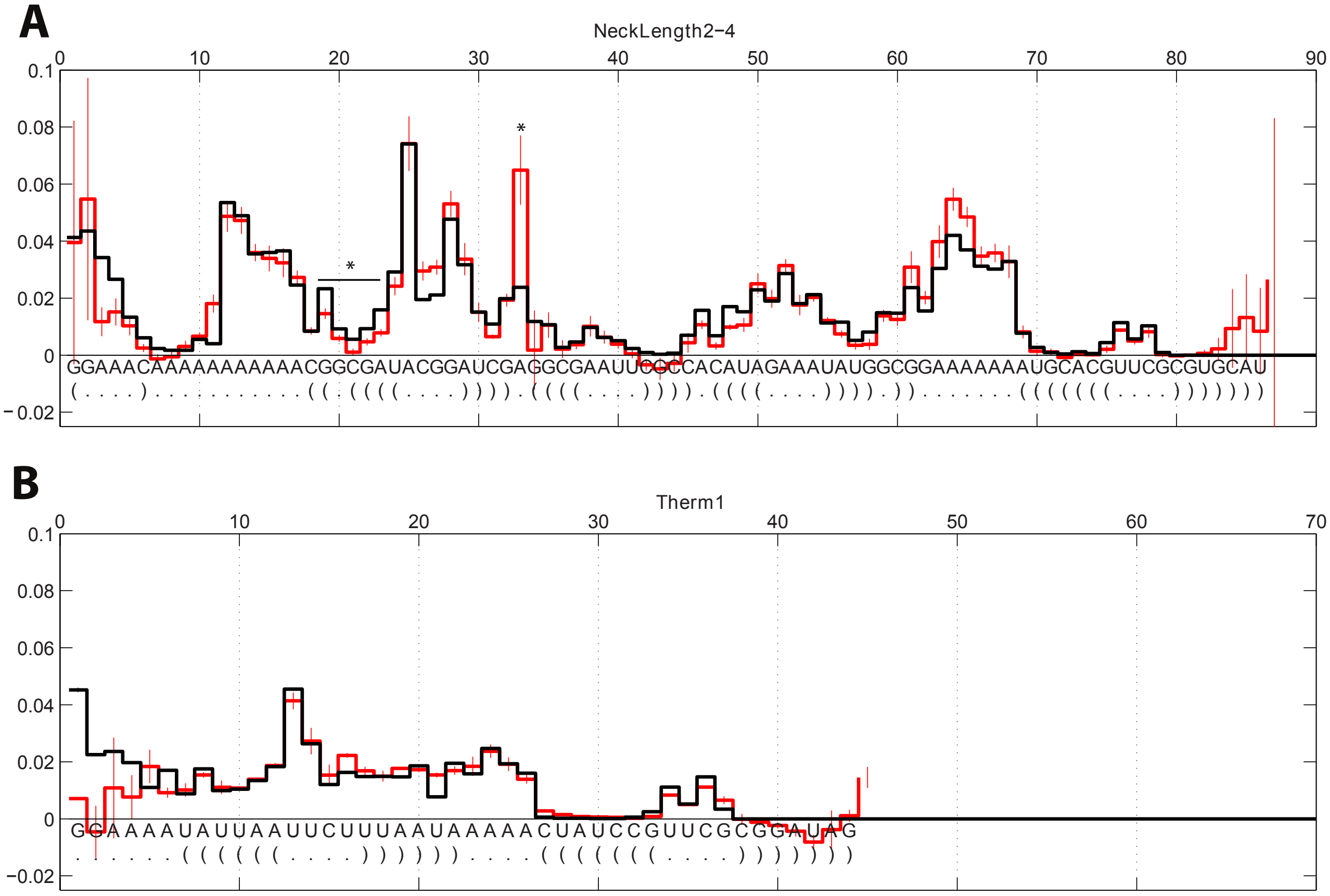

**Figure 3.** Quantitative agreement of MAP-seq data (black) with ‘gold standard’ capillary electrophoresis data (red) taken in replicate. (A) 2′-OH acylation data using the 1M7 (1-methyl-7-nitroisatoic anhydride) modifier for an RNA from the EteRNA massively parallel design platform (from a player’s project “Neck Length 2-4”). (B) 1M7 data for the Therm1 hairpin (see also Figure 2). Error bars for the MAP-seq data were estimated from statistical error on the number of reads (smaller than the line width at most sequences), as carried out by MAPseeker analysis. Error bars for the CE data were estimated after HiTRACE quantitation (36) as the standard deviation across five measurements using 1M7 concentrations of 0.625, 1.25, 2.5, 5, and 10.0 mg/mL, and an additional two replicates in the 10 mg/mL condition. Two regions that show discrepancy between the CE and MAP-seq data beyond this error are highlighted with stars; both cases can be traced to band compression in the CE experiments (not shown), which introduces systematic uncertainty in band deconvolution near guanosine residues.

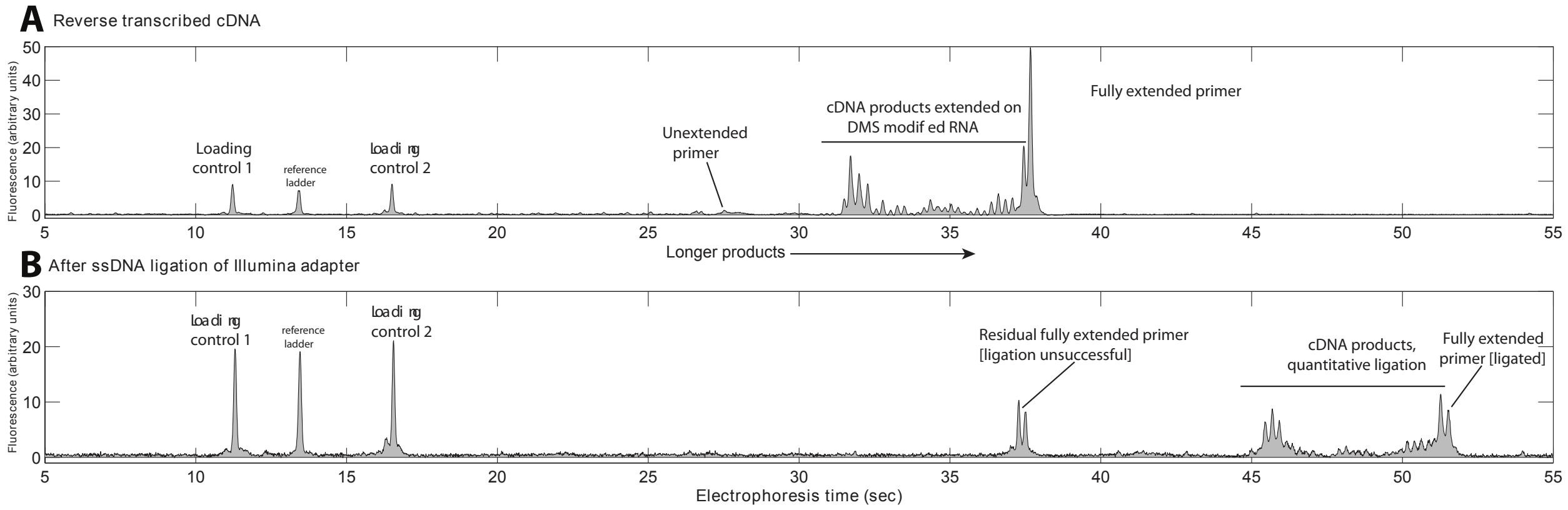

**Figure 4. Quality check and quantitation of NGS library concentration through the use of a fluorescent 'tracer' primer and capillary electrophoresis.** The sample shown corresponds to a four sample MAP-seq experiment (no modification, DMS, NMIA, CMCT) with a single RNA (Therm1; see Table 1), also shown in Figure 2. Note that the left-to-right orientation of bands is the reverse of the orders in Figures 2-3 and 5; signals corresponding to stops at more 3´nucleotides are detected at earlier times in electrophoresis. (A) The data shown are for a fluorescein-labeled primer included in the reverse transcription reaction with the DMS sample; a fraction of this reaction was loaded on an ABI 3100 sequencer sensitive to fluorescein. (B) After the final Illumina adapter ligation step. The sample loaded here on the ABI 3100 sequencer was a fraction of the sample actually loaded on the Miseq Illumina platform. A clear shift to longer electrophoresis times (longer products; right-hand peaks) is apparent. Comparison of the integrated intensity of the ligated products to loading control peaks (1 fmol fluorescein-labeled controls; left-hand peaks) gives an estimate of the total ligated product.

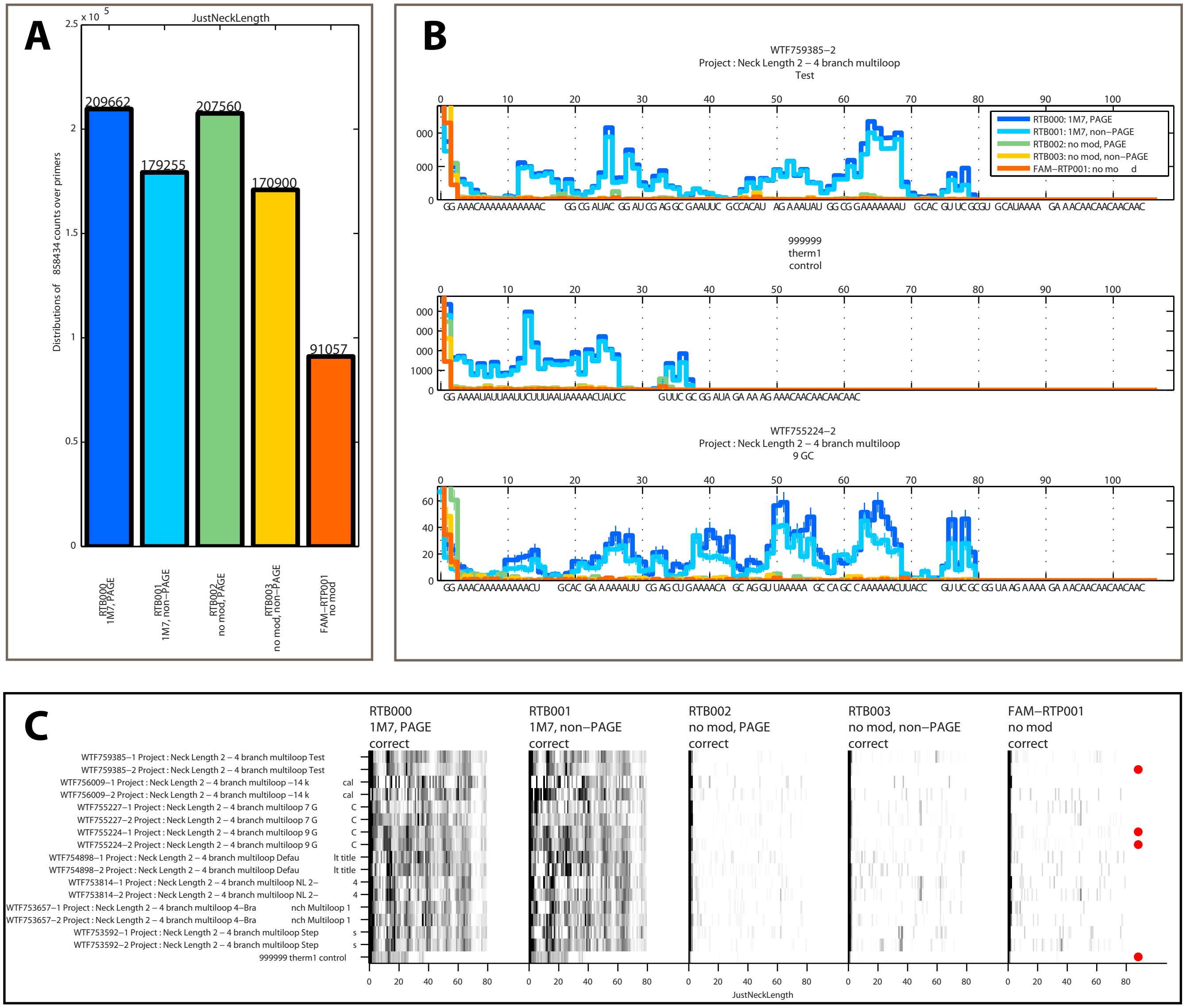

**Figure 5.** Example MAPseeker output for a MAP-seq experimental data set involving 1M7 probing of Therm1 and the Neck Length 2-4 series of RNAs (see also Figure 3). (A) Summary of number of counts across primers. In this experiment, four primers indexed two different library preparations (PAGE and non-PAGE) and modification type (1M7 or control). A fifth primer was used as a fluorescent tracer, doped into the control reactions. (B) Plot of raw counts and sequence for the RNAs with the most reads. (C) Summary display of all reactivities in heat-map format. Reactivities were automatically determined from raw counts by correcting for reverse transcription attenuation (see main text). Red symbols mark the four RNAs with the most reads.

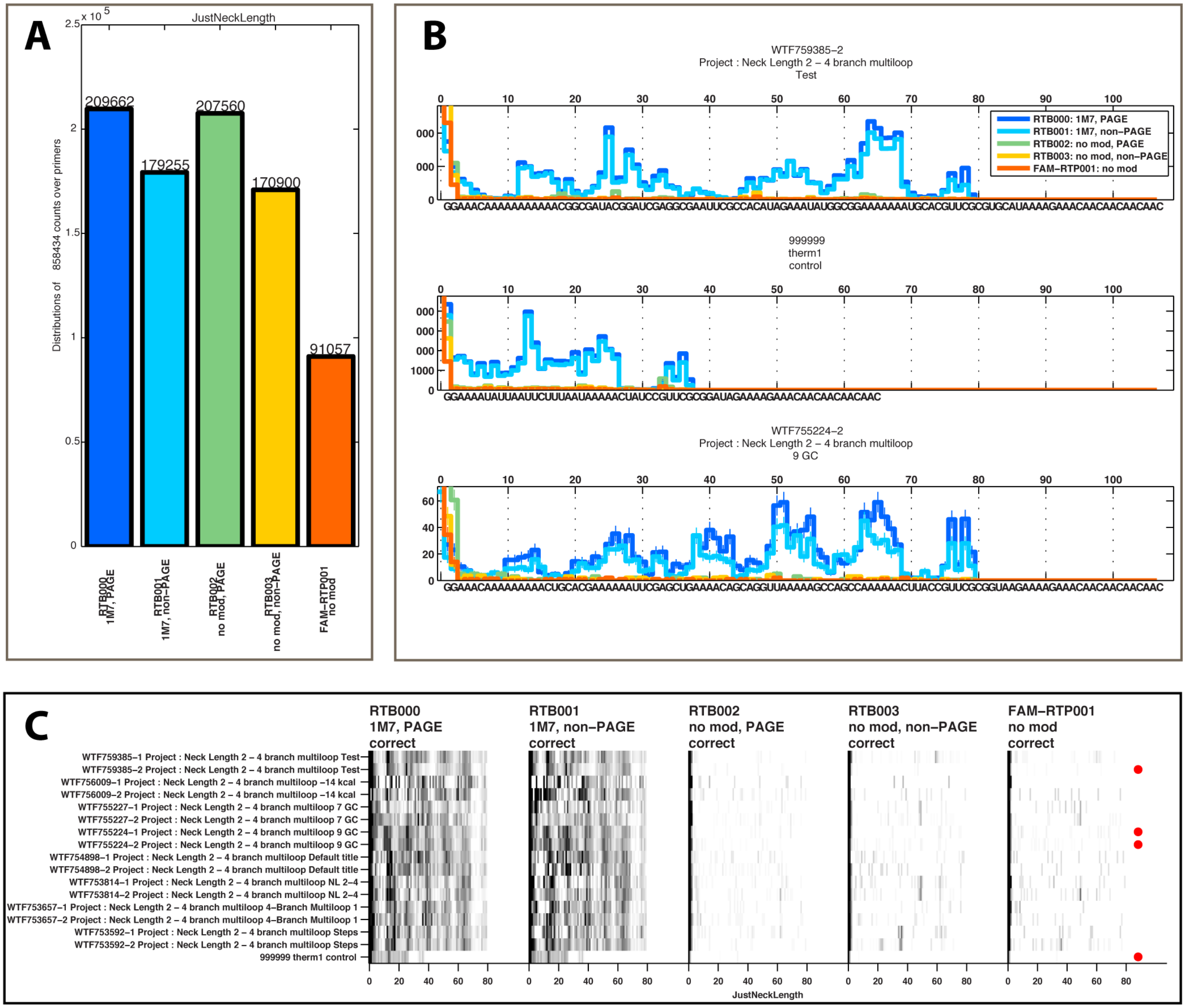

**Figure 5.** Example MAPseeker output for a MAP-seq experimental data set involving 1M7 probing of Therm1 and the Neck Length 2-4 series of RNAs (see also Figure 3). (A) Summary of number of counts across primers. In this experiment, four primers indexed two different library preparations (PAGE and non-PAGE) and modification type (1M7 or control). A fifth primer was used as a fluorescent tracer, doped into the control reactions. (B) Plot of raw counts and sequence for the RNAs with the most reads. (C) Summary display of all reactivities in heat-map format. Reactivities were automatically determined from raw counts by correcting for reverse transcription attenuation (see main text). Red symbols mark the four RNAs with the most reads.